\documentclass[letterpaper]{article}
\usepackage{natbib,alifeconf}  
\usepackage{url,hyperref,cleveref}
\usepackage{booktabs}

\usepackage{lipsum}
\usepackage{amssymb,pstricks,graphics}
\title{Fragmentation is a diversity ratchet}
\author{Russell K. Standish\\High Performance Coders}

\newcommand{\reals}{{\mathbb R}}

\newcommand{\br}{\mbox{\boldmath{$r$}}}          
\newcommand{\bbeta}{\mbox{\boldmath{$\beta$}}}   
\newcommand{\bgamma}{\mbox{\boldmath{$\gamma$}}} 
\newcommand{\bmu}{\mbox{\boldmath{$\mu$}}}       
\newcommand{\bn}{\mbox{\boldmath{$n$}}}          

\newcommand{\EcoLab}{{\sffamily\slshape
    \mbox{\raisebox{.5ex}{Eco}\hspace{-.4em}{\makebox[.5em]{L}ab}}}}

\begin{document}
\maketitle

\begin{abstract}
  A fragmented landscape reduces the impact of interspecies
  connectivity, leading to higher diversity levels than otherwise
  possible in a connected landscape. Reconnecting a previously
  fragmented landscape initiates an extinction event, preferentially
  weeding out more highly connected species. A sequence of
  fragmentation-coalescence events will drive the ecosystem to higher
  levels of diversity in a ratchet-like effect, than if the landscape
  continuously remained connected.
\end{abstract}

\section{Introduction}

250 million years ago was an absolute disaster for life on earth, with
70-80\% of species perishing within a short space of time. Since that
time, life has diversified in an exponential fashion
\cite{Benton95}. Although the end-Permian extinction event is blamed
on climate change from carbon dioxide released through a massive
volcanic event \cite{BurgessBowring15}, it is also true that 250 Mya,
the Earth's continents were drawn together in a single supercontinent
known as {\em Pangaea}, the last of many such supercontinents during
Earth's history, and that since that time, the landmass distribution
has been marked by increasing fragmentation of the continents.

The theory of {\em Island Biogeography} posits that species diversity
is proportional to $A^z$ (the {\em Species Area Relationship} (SAR)),
where $z$ is a number less than 1. Pangaea is effectively one big
island --- when the continents fragmented, the world's total diversity
increases. If an area $A$ breaks into $n$ pieces, the power law SAR
implies that diversity will grow as $n(A/n)^z \propto n^{1-z}$. This
notion that continental drift drives diversity is given the name {\em
  biogreographic provincialism}, and was studied by a number of
authors \cite{Vallentine73, Signor90, Signor94, Benton90}. I studied
this idea in the \EcoLab{} model \cite{Standish02b}, and noted that
frequent fragmentation and coalescence events appeared to cause a
ratcheting of diversity. This was a surprising result, as it was
expected that diversity should rise as fragmentation occurs, and fall
during coalescence. Instead, as reported in that paper, diversity rose
during coalescence, and remained steady during fragmentation. There
things remained --- I failed to reproduce the results in the next
version of the \EcoLab{} model, unsure whether the effect was a real
artifact of the model, or an implementation error.

However, the notion that repeated cycling of fragmentation and
coalescence, or in \EcoLab{}'s case, cycling the migration rate
between high and low values, would have a ratcheting effect on
diversity never went away. The next section describes theoretical
reasons why this ratcheting effect should be expected. 

\section{The ratchet}

\subsection{Stability and Persistence Criteria}

\cite{May72} devised a stability criteria for Lotka-Volterra
systems (of which \EcoLab{} is an example) based on linear stability
analysis about the interior equilibrium point. Put simply, for randomly
chosen connection strengths in the foodweb, stability of the interior
equilibrium point is only guaranteed if:
\begin{equation}\label{May}
  d<1/\sigma^2
\end{equation}
where $d$ is the ecosystem diversity and $\sigma$ is the standard
deviation of the off-diagonal terms of the connection matrix (with the
diagonal terms normalised to unity).

May originally developed his argument as a counter-argument to a naive
``complex ecosystems are more stable'' attitude prevalent in
ecology. Of course, May's argument applies to randomly assembled
ecosystems, and real ecosystems evolve over millenia and as such are
decidedly not ranom. Also, linear stability is the wrong criterion, as
real ecosystems tend to have an unstable equilibrium, for example with
Lotka-Volterra predator-prey systems, where predator and prey numbers
oscillate out of phase with each other. What really matters is the
concept of {\em persistence}, which is where no species becomes
extinct in the system.

For generalised Lotka-Volterra systems, on which \EcoLab{} is based,
there is a sufficiency persistence result due to
\cite{Jansen87}. There is one interior equilibrium point, plus
equilibria points on the boundaries of the positive cone $\reals^n_+$ for
sub-communities where one or more species is not present. Jansen's
condition effectively requires that each boundary equilibrium must be
unstable in the direction of the interior equilibrium point. As
discussed in \cite{Standish98b}, the probability of a randomly
assembled ecosystem persisting decreases exponentially with
diversity. As diversity builds up through speciation, eventually the
system will fail to be persistent, and an extinction avalanche will
occur until the system becomes persistent. This is the mechanism of {\em
  self-organised criticality}, modelled by Bak et al as a sand pile, where
the continuous addition of sand to the pile eventually causes the
angle of repose to be exceeded by the pile, initiating an avalanche of
sand restoring the sand pile to have slope less than the angle of
repose \cite{Bak-etal88}.

What this does mean is that evolution will tend to drive the system
towards the critical manifold, which will have a hyperbolic
relationship between $d$ and $\sigma^2$. For typical \EcoLab{} runs,
$d\sigma^2$ tends to fluctuate in the range 3.5-4, growing slowly in time.

\subsection{Species Area Relationship}

The theory of {\em island biogeography} posits a power law
relationship between the area of islands and their diversity of climax
ecosystems. In short:
\begin{equation}
  d \propto A^z.
\end{equation}

One way of understanding this is to consider the area to be made of
$A$ cells (literally the case with the cellular \EcoLab{}
model \cite{Standish98c}). If migration between cells were inhibited,
eventually the species in each cell will diverge, and overall
diversity will be proportion to the number of cells, ie $z=1$. If
migration was effectively infinite, then the system is equivalent to
the {\em panmictic} case, and sustained diversity would be independent
of area, ie $z=0$. In between the two situations, we'd expect $0<z<1$
for finite values of migration. 

\subsection{Effect of alternating fragmentation and coalescence}

When no migration occurs, each cell will evolve independently of
the others, so we'd expect diversity to grow over time. When migration
is switched on again, we expect a mass extinction to occur (such as
the {\em Great American Biotic Interchange}, which occurred when the
Panama isthmus rose out of the sea, connecting the North and South
American continents \cite{Carrillo-et20}). This will preemptively clear out the most connected
species, leading to a much reduced connectivity. Diversity will
rebound as evolution replaces the lost species, but because the
foodweb has been thinned out during the mass extinction, the rebound
will be to a higher level than the previous cycle.

In this experiment, using an artificial evolving ecology model called
\EcoLab{} \cite{Standish94} we alternate between a period of high
migration and no migration, with period of 500,000 timesteps for each
phase, which is a somewhat extreme driving force. In \cite{Standish02b}
a similar experiment was performed with the the driving force varied
in a sinusoidal fashion.

\section{Methods}

The \EcoLab{} model is based on a generalised Lotka-Volterra equation
with mutation and migration:
\begin{equation}
\dot{\bn} = \br*\bn + \bn*\bbeta\bn + {\tt mutate}(\bmu,\br,\bn) + \bgamma*\nabla^2\bn.
\end{equation}
\bn\ is the population density vector, \br\ the growth rates (net
births-deaths in absence of competition), \bbeta\ the interaction
matrix, \bmu\ the (species specific) mutation rates and \bgamma\ the
migration rate. * denotes elementwise multiplication of vectors. In the panmictic case, the \bgamma\ term is left out,
and \bn\ refers to total populations, rather than population
densities.

The mutation operator randomly adds new species $i$ into the system with
phenotypic parameters ($r_i$, $\beta_{ij}$, $\mu_i$ and $\gamma_i$)
varied randomly from their parent species.


The \bn{} vector has integral valued components --- in assigning a
real valued vector to it, the values are rounded up {\em randomly}
with probability equal to the fractional part. For instance, the value
2.3 has a 30\% probability of being rounded up to 3, and a 70\%
probability of being rounded down. Negative values are converted to
zero. If a species population falls to zero, it is considered extinct,
and is removed from the system. It should be pointed out that this is
a distinctly different mechanism than the threshold method usually
employed to determine extinction, but is believed to be largely
equivalent.

Spatial \EcoLab{} is implemented as a spatial grid, with the
$\nabla^2$ term being replaced by the usual 5-point stencil. Some care
must be taken to avoid more individuals migrating from a cell than is
contained in the cell itself. The solution found here is to cap
migration between any two cells by the neighbourhood size, so that
number of individuals travelling from an interior cell A to B is less
than a quarter of the number of individuals in cell A.

For the purposes of this experiment, we fix the value of migration
$\gamma$ to have the same value for all species, rather than being
allowed to vary through mutation and speciation. the reason being that
in setting the migration values to zero, new species will have
migration values mutated from zero, leading the misleading results
when the original migration vector is restored.

Similarly, for the mapping the surface of diversity-connectivity and
species-area relationships, both mutation and migration rates are fixed
to a single global value, to avoid confounding effects.

The source code for this experiment is available in the \EcoLab{}
package\footnote{https://github.com/highperformancecoder/ecolab,
version 6.1}. Supplementary data from the runs is available from an
Open Science Framework project \cite{Standish26b}.

\section{Results}

\subsection{Diversity-Connectivity relationship}

\begin{figure}
  \resizebox{.48\textwidth}{!}{\includegraphics{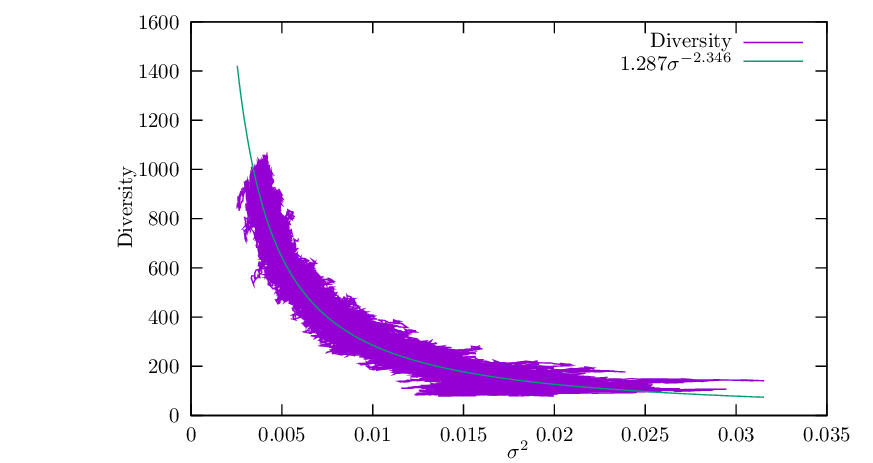}}
  \caption{Diversity as a function of connectivity ($\sigma^2$) for a
    single panmictic model run of 10 million timesteps}
  \label{diversity-connectivity}
\end{figure}

\begin{figure}
  \resizebox{.48\textwidth}{!}{\includegraphics{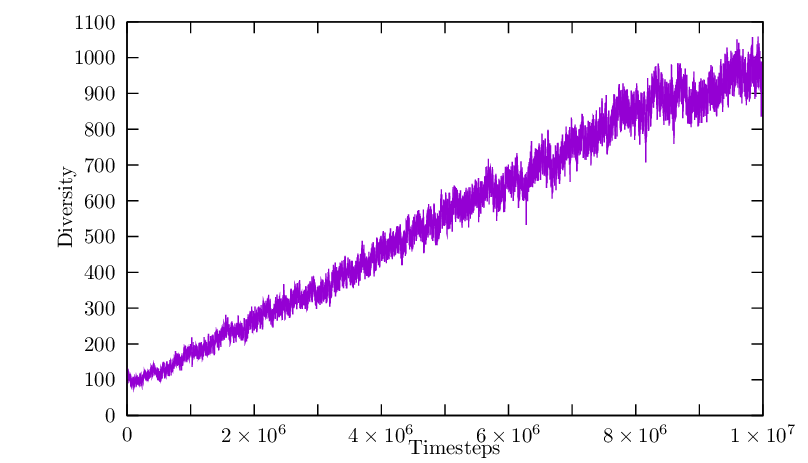}}
  \caption{Diversity as a function of time for the
    panmictic model run in Figure \ref{diversity}}
  \label{diversity}
\end{figure}

Figure \ref{diversity-connectivity} shows the trajectory of a single \EcoLab{}
run over 10 million timesteps for the panmictic model, and is
indicative of multiple such runs. The system starts with 100 species, so
in the bottom right hand corner of the plot, and moves towards the top
left. The growth in diversity is almost linear in time, as shown in
Figure \ref{diversity}.

The fitted curve was fitted using linear least squares fit on
logarithms of the data, with correlation coefficient
$\rho^2=0.999429$.

These runs are done with the \verb+gen_bias+ ($g$) parameter set to 0,
ie not biasing mutation towards either generalisation nor
specialisation. The evolutionary dynamics naturally have a
specialisation trend, which make sense in that extinction events will
preferentially target well-connected members of the ecosystem over
highly specialised members connected to just a few other species. The
question naturally arises as to why this trend was never seen before
without setting $g$ to highly negative values \cite{Standish02b}. One
possible answer is that it takes time for the increasing diversity
trend to be evident in the model run --- more than a million
timesteps. This 7 million timestep run took only a few hours on a
modern laptop, it would have been a ``hero run'' on a supercomputing
facility in 2002, and perhaps such long runs with with $g=0$ were
never performed.

The fitted curve of $D\propto \sigma^{-2.346}$ indicates the critical
manifold is not a pure hyperbolic relation, as predicted by May, but
has an additional boost due to persistence being the relevant factor,
not stability.

\subsection{Species-Area relationship}

\EcoLab{} was run as a parameter sweep for $\gamma=0,10^{-5},10^{-4}$
and $10^{-3}$, on grids $1\times1, 1\times2, 2\times2, 2\times3,
3\times3, 3\times 4$ and $4\times4$. Mutation was set at a fixed value
of $10^{-3}$, with each instance running for 10 million steps,
discarding the first 100,000 steps.

At the time of writing, not all runs had completed, so we report on
the partial dataset here.

\begin{figure}
  \resizebox{.48\textwidth}{!}{\includegraphics{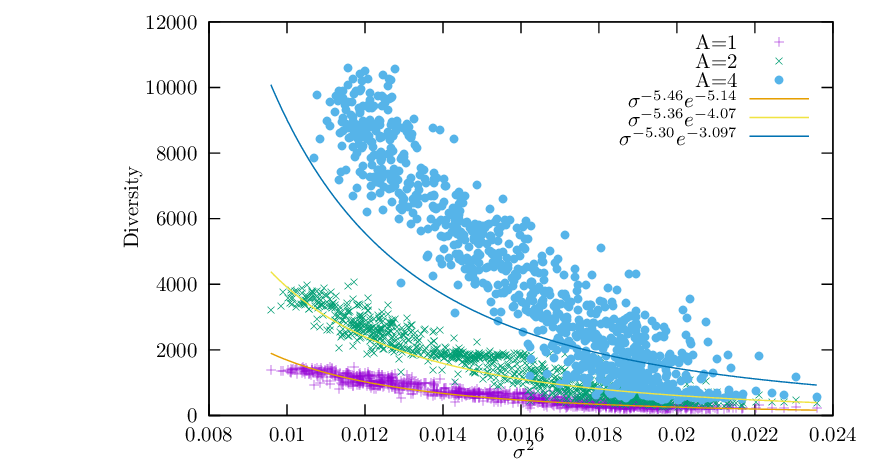}}
  \caption{Diversity-connectivity plots for runs with $\gamma=10^{-3}$
  with areas 1, 2 and 4 cells. The fitted curves are by linear
  regression of the logarithms of the data points.}
  \label{SAR}
\end{figure}

Figure \ref{SAR} shows diversity versus connectivity ($\sigma^2$) as a
function of area for $\gamma=10^{-3}$, illustrating the {\em species-area relationship},
for different sizes of the environment out to 10 million
timesteps. The fitted curve is obtained by standard linear regression
on the log-log plot. The Pearson coefficients for the fitted curves
are $-0.94$ for $A=1$, $-0.97$ for $A=2$ and $-0.99$ for $A=4$.

It should be noted there is a factor of two discrepancy in the
computed exponent between Figures \ref{diversity-connectivity} and
\ref{SAR}. At the time of writing, this hasn't been resolved.

\begin{figure}
  \resizebox{.48\textwidth}{!}{\includegraphics{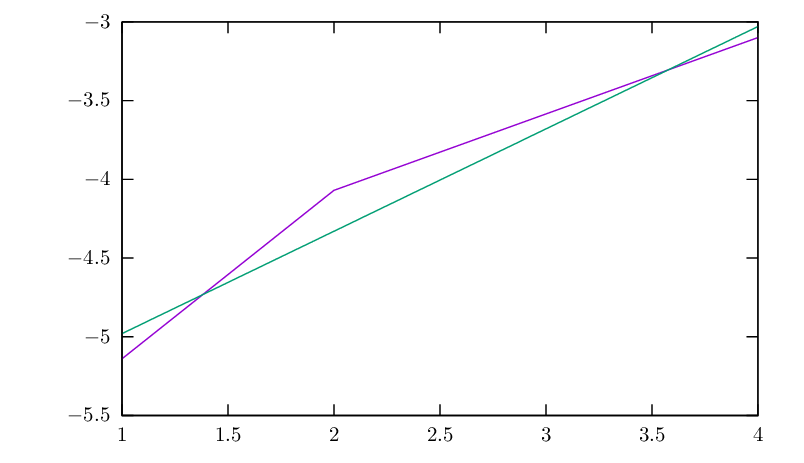}}
  \caption{Intercepts of the fitted lines on the log-log plots, as a
    function of $A$, with the fitted line indicating $z=0.65$}
  \label{intercepts}
\end{figure}

If diversity is factorisable as
\begin{equation}\label{SAR-factorisable}
  d \propto \sigma^y A^z
\end{equation}
then we'd expect the slopes of the above fits to be independent of
area, and the intercepts of the above fits on the log-log data should
follow a linear relationship with slope $z$. Figure \ref{intercepts}
shows the intercept data for $\gamma=10^{-3}$, and the fitted line for $z=0.65$,
which of all the migration cases bears the strongest similarity to eq
\ref{SAR-factorisable}.

For lower values of $\gamma$, the diversity-connectivity exponent has
a substantial dependence on area, for reasons that are not clear at
present. The full dataset is available in the supplementary materials,
and will be updated as the runs finish.

\subsection{Fragmentation Ratchet}

\begin{figure}
  \resizebox{.48\textwidth}{!}{\includegraphics{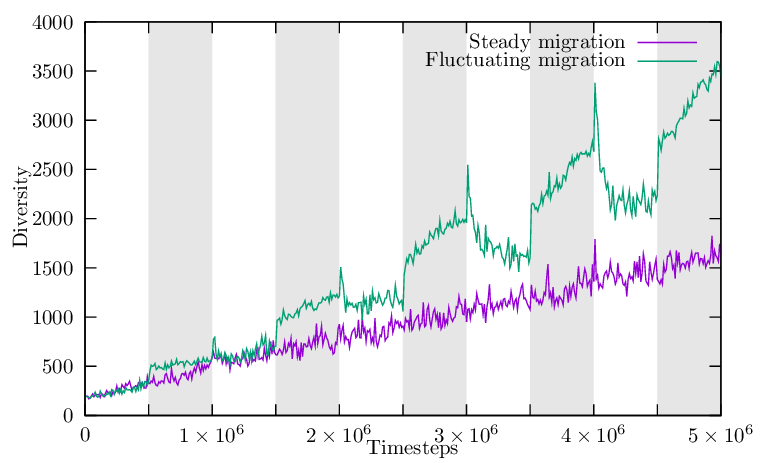}}
  \caption{Diversity growth on a $2\times2$ grid under steady migration
    $\gamma=10^{-3}$ or fluctuating between $10^{-3}$ and zero every
    500,000 timesteps. Grey patches indicate when migration drops to
    zero in the fluctuating case.}
  \label{fragmentation-pump}
\end{figure}

In Figure \ref{fragmentation-pump}, we see the ratchet in action. When
migration drops to zero, such as when a continent fragments, or sea
level rises to create a set of islands from what were previously
mountaintops, evolution speeds up the creation of new species. When
migration returns, eg when North and South America collided, or in the
present day when global shipping transports species around the world,
there is an extinction event, dropping diversity to the critical curve
for the current ecosystem connectivity. Such extinction events
preferentially remove highly connected species, leading to a gradual
walk along the critical manifold towards high diversity. Over time, the
sequence of fragmentation and coalescence events and associated
extinction events. cause diversity to grow faster than if migration
levels remained constant through time, with its smaller, more steady
extinctions.

\section{Discussion}

When looking at the fossil record \cite{Benton95}, we do see a period of
saturation (the ``paleozoic plateau'') leading up the end-Permian extinction,
which coincides with the break up of Pangaea, followed by a rebound
that exhibits an almost exponential growth in diversity. It is
circumstantial evidence at best, but evidence supporting the
fragmentation ratchet hypothesis.

Another more local example of fragmentation and coalescence occurs
when mountaintops become islands as sea levels rise as glaciers
retreat during interglacial period - conversely glaciations lock up
sea water, leading to islands becoming part of the mainland. This
should have an impact on the evolutionary process of an
archipeligo. It would be interesting to see a study of diversity
trends of an archipeligo over the Quaternary (2.6Mya--present), which
has exhibited significant cycling between glacial and interglacial
periods.

It might seem that this work has implications for environmental
conservation. That ecosystem fragmentation is perhaps not the
boogeyman of conservation, assuming biodiversity is the goal of
conservation. However, the phenomena idenitifed in this paper plays
out over evolutionary timescales, and conservation goals are usually over
human timescales. It is well known that fragmentation can promote
extinction of vulnerable species, and evolution's replenishment
process is too slow to be relevant.

Whilst \EcoLab{} is perhaps the first evolutionary ecology model, it is
not the only one in existence, with WebWorld \cite{Drossel-etal01} and
Tangled Nature \cite{Christensen-etal02} being notable alternatives. WebWorld,
in particular, has shown no signs of exhibiting the self-organised
criticality seen in \EcoLab{}, so may well not exhibit the power law
species-area relationship either. It would be interesting to see if
this fragmentation ratchet phenomenon is present or not in that system.

\bibliographystyle{apalike}
\bibliography{rus}
\end{document}